\documentclass[12pt,preprint]{aastex}

\shorttitle{Star Cluster SNIa}
\shortauthors{Shara \& Hurley}

\slugcomment{To appear in Astrophysical Journal: accepted February 7, 2002}

\begin{document}

\title{Star Clusters as Type Ia Supernova Factories}

\author{Michael M. Shara and Jarrod R. Hurley}
\affil{Department of Astrophysics,
       American Museum of Natural History,
       Central Park West at 79th Street, \\
       New York, NY 10024}
\email{mshara@amnh.org, jhurley@amnh.org}

\begin{abstract}
We find a remarkably enhanced production rate in star clusters
(relative to the field) of very short period, massive double-white-dwarf 
stars and of giant-white dwarf binaries.
These results are based on $N$-body simulations performed with the new 
GRAPE-6 special purpose hardware and are important in identifying and 
characterizing the progenitors of type Ia supernovae. 
The high incidence of very close double-white-dwarf systems is the result 
of dynamical encounters between (mostly) primordial binaries and other 
cluster stars.
Orbital hardening rapidly drives these degenerate binaries to
periods under $\sim 10$ hours.
Gravitational radiation emission and mergers producing supra-Chandrasekhar 
objects follow in less than a Hubble time.
If most stars are born in clusters then estimates of the double white dwarf 
merger rates in galaxies (due to cluster dynamical interaction) must be 
increased more than tenfold.
A majority of the Roche lobe overflow giant-white dwarf binaries are not 
primordial; they are produced in exchange reactions.
Most cases resulted in a common-envelope and formation of a double-white-dwarf 
binary rather than Supersoft X-ray sources leading possibly to a 
type Ia supernova.
\end{abstract}


\section{Introduction}
\label{s:intro}

Type Ia supernovae (SNeIa) have recently been used to demonstrate that the 
Universe is apparently not only expanding, but also accelerating 
\citep{rie98,per99}.
If this remarkable discovery survives careful scrutiny it has profound 
implications for cosmology physics (see Leibundgut 2001 for a review). 
However, before accepting such a revolutionary
change in our view of the universe, it is critical to investigate
all challenges to the acceleration interpretation of the data.
A thorough compilation of those challenges is given by
\citet{rie00}.

\citet{rie00} cites evolution, dust, gravitational lensing, measurement 
biases, selection biases and alternative cosmological models as potential 
challenges to the acceleration interpretation
of the Type Ia supernova (SNIa) observations. He concludes that
the primary source of reasonable doubt is evolution
(could SNeIa at redshift z = 0.5 be intrinsically fainter than
nearby SNeIa by 25\% ?). We simply don't know
for certain what kind of star (or stars?) give rise to
these explosions, and hence whether to expect systematic variations in 
SNIa luminosities with z.

Not knowing the progenitors of SNeIa is embarrassing because of the 
significant empirical corrections one must apply to supernova
luminosities, based on their light curves, to get distances \citep{phi93}. 
If one could determine the SNIa progenitor with a high degree of reliability, 
one could build increasingly sophisticated supernova light curve models. 
This would lead to a fundamental understanding of the physical behavior of
one of the most crucial standard candles in cosmology.

Two competing SNIa models exist: merging double-white-dwarfs (DWDs) and 
accreting single white dwarfs (ASDs) in close binary
systems (see Yungelson \& Livio 2000 for a review, 
including the pros and cons of each flavor of each model). 
In both cases the model involves the thermonuclear disruption of a 
white dwarf, most likely of carbon-oxygen (CO) composition, 
when its mass reaches, or exceeds, the critical Chandrasekhar mass. 
However, \citet{sai98} have raised the possibility that supra-Chandrasekhar 
mass-accreting white dwarfs will undergo accretion-induced collapse (AIC) 
and the formation of a neutron star, rather than a SNIa. 
An important example of why it is so critical to know the progenitors of 
SNeIa is the following. ASDs accreting helium result in the accumulation of a 
He layer and an edge lit detonation; the peak luminosity-light curve shape 
relation of such objects may vary significantly with metallicity and hence 
redshift \citep{tou01}.
 
It is clearly important to calculate the expected incidence of DWDs and ASDs 
in the stellar populations of the different types of galaxies where SNIa occur. 
This has been done for field stars \citep[for example]{yun96}, but hardly for 
the environs of clusters. 
We have begun this study, and find that dynamics dramatically alter binary 
populations and characteristics, including those of DWDs and ASDs.
Such an effect has been predicted in the past \citep{che93} but has 
yet to be tested by direct means. 

Many, and possibly all, stars were born in a star cluster \citep{kra83,lad93}, 
or at the very least a loose association \citep[for example]{mfa01}. 
Living in an environment with stellar densities of 
$10^2\,$stars$\,{\rm pc}^{-3}$ to as high as $10^7\,$stars$\,{\rm pc}^{-3}$ 
can dramatically affect the evolution of stars. 
Physical collisions, and, in the case of binaries, exchange interactions or 
disruptions of orbits can radically alter the fates of cluster stars. 
The most direct way to model the
evolution of a star cluster is with an $N$-body code in which the individual 
orbits of each star are followed in detail {\it and} the internal evolution 
of each star is also taken into account \citep{hur01}.

We have begun a study of the behavior of populous star clusters using a 
state-of-the-art $N$-body code in conjunction with the powerful GRAPE-6 
special purpose computer \citep{mak01}.
This will ultimately involve a large number of
$N$-body simulations covering a wide range of initial conditions,
e.g. metallicity, binary fraction, and stellar number density.
A related study investigating the fate of planetary systems in star 
clusters is also ongoing \citep{hsh02}. 
The remarkable evolution of close binaries comprised of one or two 
degenerate members is the focus of this early work.
The increased importance of SNIa for cosmology makes these initial results 
particularly relevant to the scientific community. 

Our simulation method is presented in Section 2, and the 
double white dwarfs are described in Section 3. 
The accreting white dwarfs are detailed in Section 4. 
We briefly summarize our results in Section 5. 

\section{Simulation Method}
\label{s:method}

We have carried out simulations with $22\,000$ stars and a 10\% primordial 
binary fraction using the Aarseth {\tt NBODY4} code \citep{aar99,hur01}. 
A GRAPE-6 board located at the American Museum of Natural History hosts 
the code. 
This special purpose computer acts as a Newtonian force accelerator for 
$N$-body calculations, performing at $0.5\,$Tflops for a 16-chip board 
($\sim 30\,$Gflops per chip).

The prescription used for single star evolution in {\tt NBODY4} is described 
in \citet{hur00}. 
A feature of this algorithm is the inclusion of metallicity as a free 
parameter, making it applicable to modelling clusters of all ages. 
It covers all stages of the evolution from the zero-age main-sequence (ZAMS) up to, 
and including, remnant stages such as the white dwarf cooling track. 
In terms of examining DWDs as SNIa candidates it is important to note that 
the WD initial-final mass relation found by \citet[see their Fig.~18]{hur00} 
is well matched to observations. 

All aspects of standard binary evolution, i.e. non-perturbed orbits, 
are treated according to the prescription described in \citet{hur02}. 
This includes tidal circularization and synchronization of the orbit, 
mass transfer, and angular momentum loss mechanisms such as 
magnetic braking and gravitational radiation. 
The default input parameters to this algorithm 
(listed in Table~3 of Hurley, Tout \& Pols 2002) are adopted.
In particular, this means that the common-envelope (CE) efficiency parameter 
is taken to be ${\alpha}_{\rm CE} = 3.0$ which makes the outcome of 
common-envelope evolution similar to the alternative, and commonly used, 
scenario described by \citet{ibe93} with ${\alpha}_{\rm CE} = 1.0$.
Common-envelope evolution occurs when mass transfer develops on a dynamical 
timescale. 
In the evolution algorithm this equates to the donor star, generally 
a giant, having an appreciable convective envelope and a mass-ratio, $q$, 
exceeding some critical value, $q_{\rm crit} \simeq 0.7$. 
If the conditions for dynamical mass-transfer are met then 
the envelope of the giant overfills the Roche-lobes of both stars leaving 
the giant core and the secondary star contained within a common-envelope. 
Owing to orbital friction these will spiral together and transfer energy 
to the envelope with an efficiency ${\alpha}_{\rm CE}$. 
If this process releases sufficient energy to drive off the entire envelope 
the outcome will be a close binary consisting of the giant core and 
the secondary, otherwise it leads to coalescence of the two objects. 

Various models for binary evolution exist
\citep[for example]{por96,tut96,tou97}
with each having the similar goal of providing a sufficiently detailed description
of binary behaviour while remaining computationally efficient.
The main difference between these and the \citet{hur02} model referred to in this
work, and incorporated in to {\tt NBODY4}, is the much improved treatment of
tidal interactions present in the latter.
Tidal friction arising from convective, radiative or degenerate damping mechanisms
is modelled and necessarily the stellar spins, which are subject to tidal
circularization and synchronization, are followed for each star.
Also, by using the single star evolution algorithm of \citet{hur00} the
\citet{hur02} binary prescription not only allows for a wider range of evolution phases,
with many of these modelled in more detail and based on updated stellar models,
but can also be used to evolve binaries of any metallicity.
An additional difference is that models such as those of \citet{por96} and \citet{tut96}
follow the \citet{ibe93} common-envelope scenario while \citet{hur02} utilise the
scenario first described in \citet{tou97}. 
Subtle variations also exist from model to model in the way that various aspects of
mass transfer are dealt with.

Gravitational radiation is an important process in the evolution of close
binary systems because it provides a mechanism for removing angular
momentum and driving the system towards a mass transfer state, possibly
followed by coalescence.
In the \citet{hur02} binary model orbital changes due to gravitational radiation
are calculated using expressions based on the weak-field approximation of general
relativity \citep{egg02} and assuming the stars are point masses.
This includes a strong dependence on the eccentricity of the orbit,
although DWD systems are generally circular. 

In the dense environment of a star cluster it is possible for the
orbital parameters of a binary to be significantly perturbed owing
to close encounters with nearby stars. 
It is even possible for the orbit to become chaotic as a result of such an 
interaction. 
Modelling of such events has been considered in detail by \citet{mar01} 
whose work is included in NBODY4.
Three-body and higher-order subsystems are also followed in detail 
\citep[and references within]{aar99}. 

We include the outcome of four simulations in the results 
described below and in Tables 1-3.
The first two simulations were carried out assuming a metallicity of 
$Z = 0.004$, while the third and fourth simulations had $Z = 0.02$. 
In all other respects the initial conditions were identical for each 
simulation. 
Masses for single stars are chosen from the initial mass function (IMF) of 
\citet{kro93} within the limits of $0.1 - 50 M_\odot$. 
For primordial binaries the total mass of the binary is chosen from the IMF 
of \citet{kro91}, as this was not corrected for the effect of binaries, 
between the limits of $0.2 - 100 M_\odot$. 
The component masses are then assigned according to a uniform distribution 
of mass-ratio, taking care to ensure that the single star mass limits are 
not violated. 
Following \citet{egg89} we take the distribution of orbital separations 
for the primordial binaries to be 
\begin{equation}\label{e:sepdst}
\left( \frac{a}{a_m} \right)^{\beta} = \sec\left( kW \right) +
\tan\left( kW \right) \, ,
\end{equation}
where $W \in \left[ -1,1 \right]$, and uniformly distributed and $k$ satisfies
\begin{equation}
\sec k = \frac{1}{2} \left[ {\zeta}^{\beta} + {\zeta}^{- \beta} \right] \, .
\end{equation}
This distribution is symmetric in $\log a$ about a peak at $a_m$ and
ranges from a minimum separation of $\zeta a_m$ to a maximum of $a_m/\zeta$.
We choose the constants $\zeta$ and $\beta$ to be $10^{-3}$ and $0.33$
respectively, and take $a_m \simeq 30\,$AU, i.e. each separation, $a$, 
is within the range 
$\sim 6 \, R_{\odot}$ to $30\,000\,$AU. 
The eccentricity of each binary orbit is taken from a thermal distribution 
\citep{heg75}. 
Initial positions and velocities of the stars are assigned according to a 
Plummer model \citep{aar74} in virial equilibrium. 

Each simulation started with $18\,000$ single stars and $2\,000$ binaries and 
was evolved to an age of $4.5\,$Gyr when $\sim 25\%$ of the 
initial cluster mass remained and the binary fraction was still close to 10\%.
In real time each individual simulation took approximately five days to 
complete, which corresponds to $\sim 10^3$ cluster crossing-times, and 
$\sim 10^{17}\,$ floating-point operations on the GRAPE board. 
Clusters are evolved subject to a standard three-dimensional Galactic tidal
field (see Hurley et al. 2001) of standard, i.e. local, strength
\citep[for example]{che90}
so, in addition to mass loss from stellar evolution, mass is removed from the
cluster when stars cross the tidal boundary and are lost to the Galaxy.
Because the orbits of some bound stars may momentarily take them outside of the
cluster, stars are not actually removed from the simulation until their distance
from the cluster centre exceeds twice the tidal radius.
At any moment in time less than 2\% of the mass in a simulation is found to lie
outside of the tidal radius and this has little effect on the evolution of
the model \citep{gie97}. 
Typically the velocity dispersion of the stars in these model clusters was 
$2\,{\rm km } \,{\rm s}^{-1}$ with a core density of 
$10^3\,$stars$\,{\rm pc}^{-3}$. 
The density of stars at the half-mass radius is generally a factor of 10 
less than this.
These simulations are clearly in the open cluster regime. 

\section{Double White Dwarfs}
\label{s:result1}

Double white dwarf systems must form in all stellar populations with binaries. 
To form short-period systems isolated binaries must undergo considerable 
common envelope evolution to shed 
angular momentum and bring their degenerate components close together. 
Only when the orbital period of a DWD is less than about 10 hours will 
gravitational radiation force the system to merge in less than the age of the 
universe. 
{\it A key result of this paper is that DWDs formed in
clusters can be significantly hardened by interactions with passing stars. 
Alternatively, hardening of binaries before they reach the DWD stage, and/or 
exchange interactions, 
can produce a short-period DWD that would not have formed in isolation. 
This greatly increases the number of DWDs which merge in less than a Hubble time}. 
Heggie (1975) defined hard binaries to be sufficiently close that their 
binding energy exceeds the mean kinetic energy of the cluster stars. 
A binary is said to {\it harden} when an interaction with a third body 
removes energy from the orbit and thus reduces the separation. 
Three-body interactions may also lead to an exchange in which one of the 
binary components is displaced by the incoming third star. 
If an exchange does occur then the expelled star, generally the least 
massive, invariably leaves the three-body system altogether. 
Note that four-body interactions involving a second binary are also possible. 
The likelihood of a binary being the target of an exchange interaction 
scales linearly with the orbital separation of the binary \citep{heg96} 
and is also more likely to occur in the core of a cluster. 
Table~\ref{t:table1} lists the characteristics of the merging DWDs formed 
during the four cluster simulations: epoch of formation, 
types of white dwarf, component and system masses, orbital period at formation 
epoch, gravitational inspiral timescale, and whether the binary is 
primordial or due to an exchange. 
All of these objects will merge 
in less than a Hubble time, creating a supra-Chandrasekhar object
which may yield a SNIa.

From binary population synthesis alone we expect about 10 DWDs to be present 
amongst 2000 binaries of solar metallicity after $4\,$Gyr of evolution. 
This expected number rises to 15 DWDs for $Z = 0.004$ which is mainly a 
reflection of the accelerated evolution timescale for low metallicity stars 
with masses less than $\sim 9 M_\odot$: the main-sequence (MS) lifetime for 
stars of Population II composition is roughly a factor of two shorter than for 
stars of solar composition (see Fig.~4 of Hurley, Pols \& Tout 2000). 
Of these DWDs, only $\sim 0.2 - 0.3$ are expected to be a ``loaded gun'', 
i.e. to have a combined mass, $M_{\rm b}$, greater than the Chandrasekhar mass, 
$M_{\rm Ch} \simeq 1.44$, {\it and} a merger timescale less than the age of  
the Galaxy \citep{hur02}. 
These population synthesis results refer to {\it isolated} binaries with 
initial conditions drawn from the same distributions as used in the $N$-body 
simulations presented here. 
From Table~\ref{t:table1} we see that there are, on average, four ``loaded guns''  
per 2000 binaries, which is $\sim 15$ times more than expected in a binary 
population unaffected by dynamical interactions. 
We note CO-CO DWDs dominate, but that CO-ONe binaries are also common 
(5 of 16 binaries). 
The system total masses range from $1.49 M_{\odot}$ to $1.96 M_{\odot}$. 
This range in masses and compositions suggests diversity in
the light curves and spectra of the merging objects, whether or
not they make SNIa.  

\begin{figure}
\epsscale{.9}
\plotone{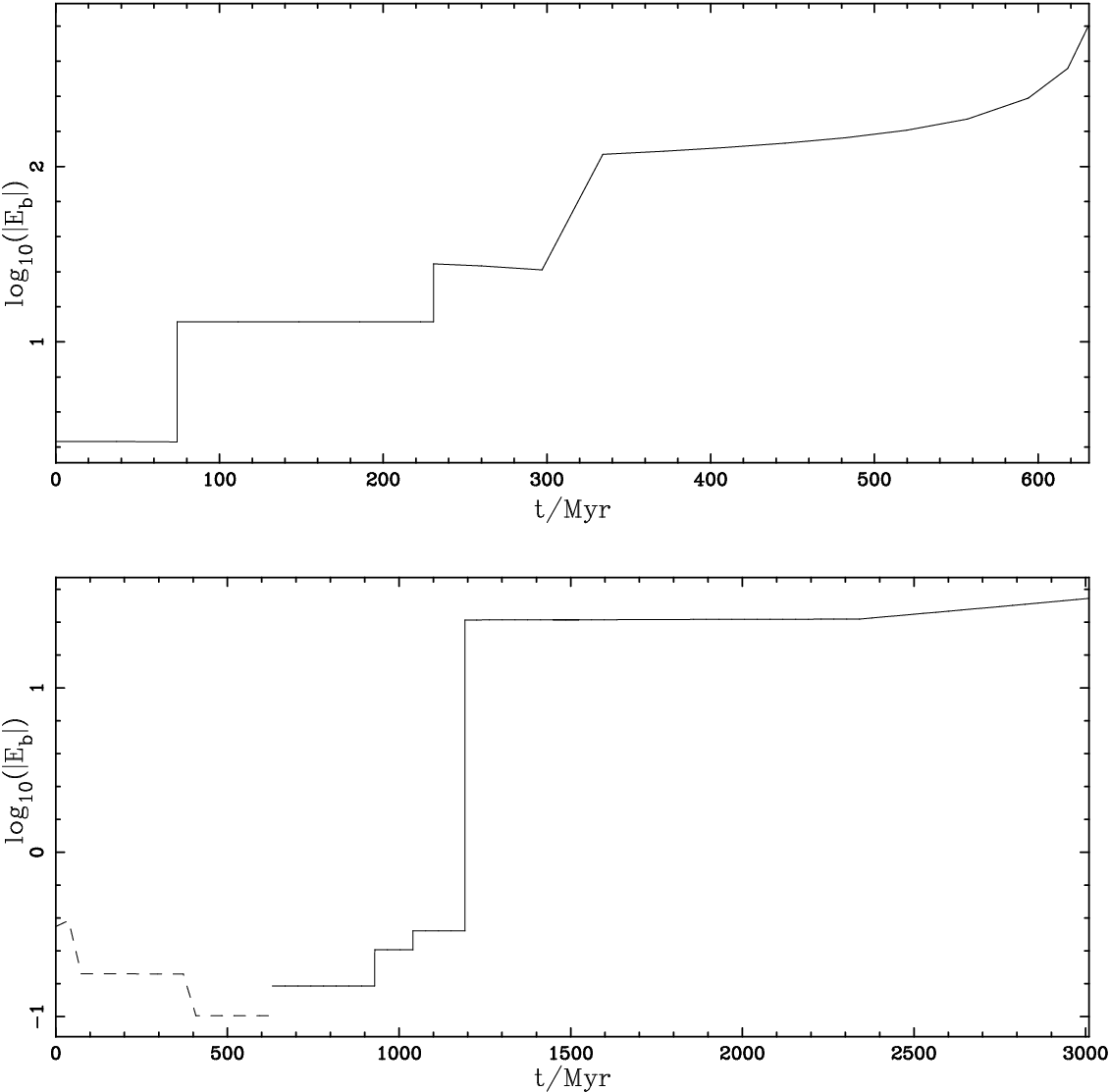}
\caption{
The evolution of particular binary systems plotted as the logarithm 
of the absolute value of the binding energy, $E_{\rm b}$, where 
$E_{\rm b}$ is in units of $M^2_\odot / AU$. 
The top panel shows the ONe-CO DWD formed at $334\,$Myr in the first 
$Z = 0.004$ simulation. 
This system is a possible SNIa candidate (fourth entry in Table~1) and 
is a primordial binary that had its orbit perturbed after the DWD formed. 
The system ceases to exist when, at $630\,$Myr, the two WDs merge to 
form a supra-Chandrasekhar object. 
The lower panel shows the ONe-CO DWD formed at $1\,192\,$Myr in the second 
$Z = 0.02$ simulation. 
This system is also a possible SNIa candidate (last entry in Table~1) and 
formed in an exchange interaction at $620\,$Myr. 
The evolution of the primordial binary involved in the exchange is shown 
as a dashed line. 
The evolution after $3\,000\,$Myr, in which the two WDs continue to spiral 
together owing to gravitational radiation, is omitted for clarity. 
\label{f:fig1}
}
\end{figure}

Figure~\ref{f:fig1} highlights the evolution of two of the SNIa candidates 
formed in the simulations. 
The binary shown in the top panel of the figure is the DWD formed at 
$334\,$Myr in the first $Z = 0.004$ simulation (fourth entry in Table~1). 
This began as a primordial binary with component masses of $5.82 M_\odot$ 
and $3.13 M_\odot$, an eccentricity of 0.74, and an orbital period of 
$2\,138\,$d. 
It underwent two common-envelope events, the first at $76\,$Myr 
when the $5.82 M_\odot$ star filled its Roche-lobe on the asymptotic giant 
branch (by which time tides raised on the convective envelope of the
primary had circularized the orbit) 
and became an ONe WD, and the second at $232\,$Myr when the 
$3.13 M_\odot$ star filled its Roche-lobe on the first giant branch 
and became a naked helium star. 
The helium star subsequently evolved to become a CO WD at $299\,$Myr, 
losing some mass in the process which explains the decrease in binding energy. 
The DWD resided in the core of the cluster and experienced a strong 
perturbation to its orbit shortly after formation which caused the orbit 
to {\it harden}. 
The perturbation did not induce any noticeable eccentricity in to this
already circular orbit. 
Gravitational radiation then removed orbital angular momentum from the 
system, causing the two WDs to spiral together, until they merged at 
$\sim 630\,$Myr. 
Without the perturbation to its orbit the period of this DWD binary would 
not have been short enough for gravitational radiation to be efficient. 

The binary shown in the lower panel of Figure~\ref{f:fig1} is the DWD 
formed at $1\,192\,$Myr in the second of the $Z = 0.02$ simulations 
(last entry in Table~1). 
This binary is the result of a 3-body exchange interaction which occurred 
at $\sim 620\,$Myr. 
The binary involved in the exchange is a primordial binary that had 
experienced two common-envelope events and was itself a DWD at the 
time of exchange (see the second entry in Table~2 for Simulation~4). 
The incoming star had a mass of $2.02 M_\odot$ and after spending a 
short time as a quasi-stable 3-body system the least massive star, the 
$0.82 M_\odot$ WD in the primordial, was ejected from the system to 
leave a MS-WD binary with an eccentricity of 0.63 and an orbital 
period of $14\,125\,$d. 
This binary was significantly perturbed at $914\,$Myr 
(increasing the eccentricity to 0.94) 
and then experienced a series of common-envelope phases 
(with the orbital eccentricity having been removed by tidal
friction prior to the first CE event) 
before forming the SNIa candidate of interest at $1\,192\,$Myr. 
It bears repeating that possible SNIa systems of this sort never occur 
naturally in the field. 

In Table~\ref{t:table2} we list all DWDs which are {\it not} SNIa candidates. 
This is because the total system masses are less than $M_{\rm Ch}$,  
or the merger timescale is longer than a Hubble time, or both. 
The large numbers of objects demonstrate the rich variety of DWDs created 
in clusters. 
It is particularly remarkable that 93 of the 135 binaries listed in 
Table~\ref{t:table2} were created in exchange interactions. 
This highlights a second key result of our paper: {\it most of the DWDs in star 
clusters, and possibly those in the field if they were born in clusters, 
have progenitors with companions different from the ones we observe today. 
Furthermore, the orbital period and mass ratio distributions
of such objects cannot be reliably predicted without including the effects of 
dynamics and thus, exchanges.} 
This is particularly important
in comparing the results of observational searches for ``loaded guns'' 
with theoretical population predictions \citep{saf98}.
The number of DWDs formed from primordial binaries in the simulations matches 
well the predicted numbers from the non-dynamical population synthesis. 
This is also true of the SNIa candidates in Table~1 where the population 
synthesis predicted either 0 or 1 candidate per simulation. 
 
\begin{figure}
\epsscale{1.0}
\plotone{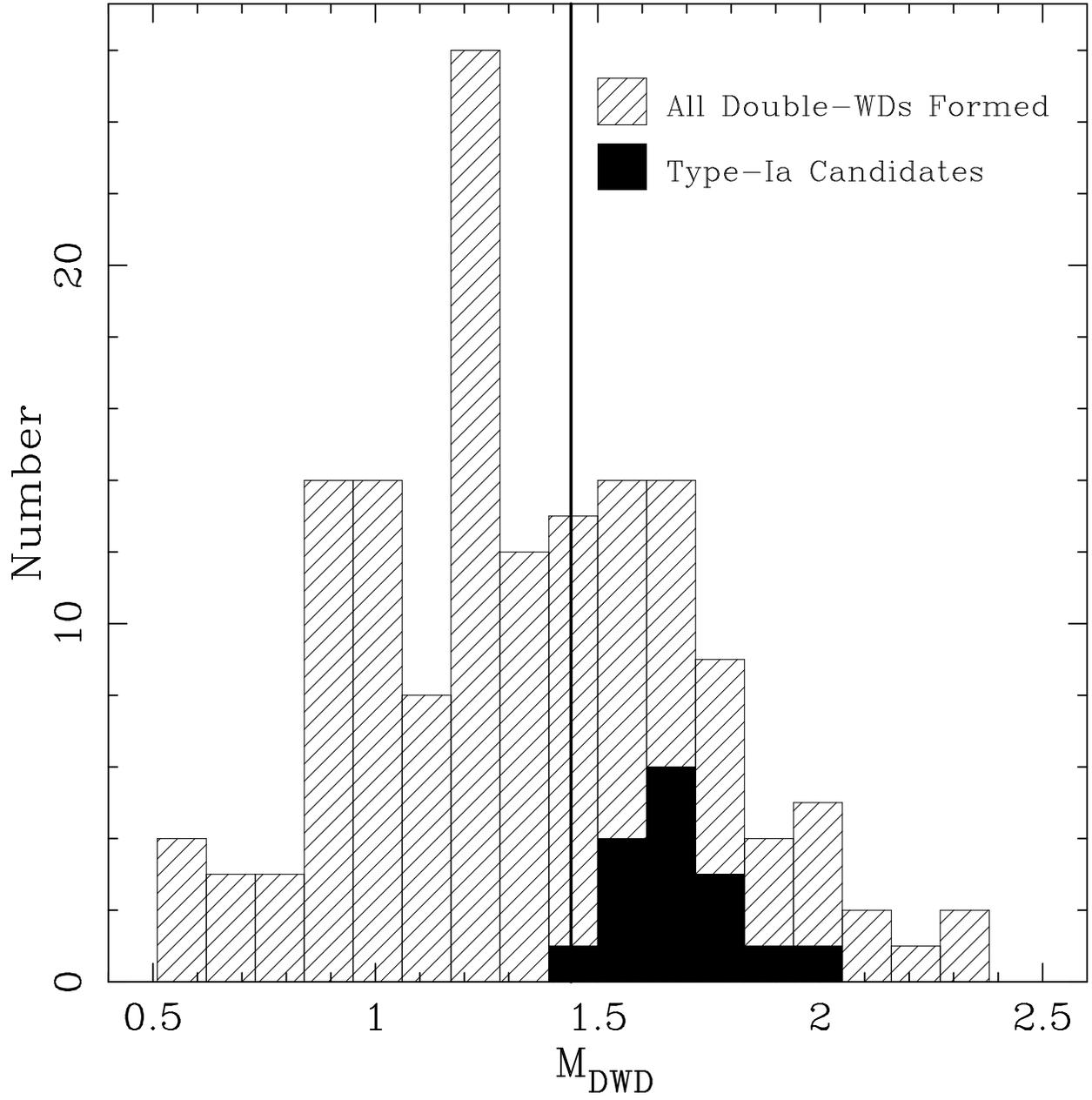}
\caption{
Histogram of double-WD masses.
The Chandrasekhar mass of $1.44 M_{\odot}$ is also shown.
All systems listed in Tables~1 and 2 are included.
\label{f:fig2}
}
\end{figure}

\begin{figure}
\plotone{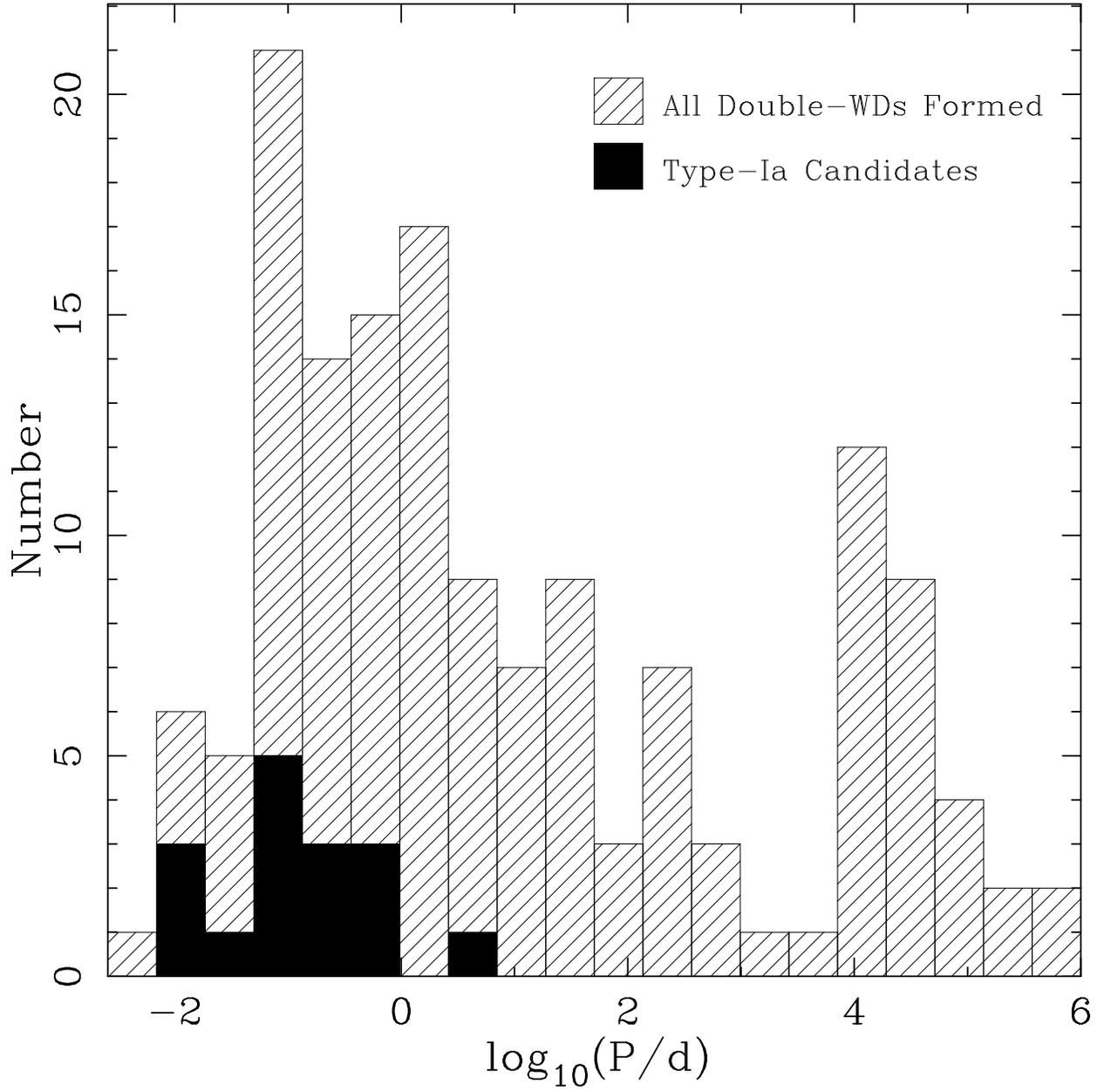}
\caption{
Histogram of double-WD periods at the time of formation.
All systems listed in Tables~1 and 2 are included.
\label{f:fig3}
}
\end{figure}

The histogram of all DWD masses is shown in Figure~\ref{f:fig2}.
The distribution of DWDs which will merge in less than a Hubble time is quite
similar to that of the DWDs with $M_{\rm b} > M_{\rm Ch}$.
This is also clear from the histogram of all DWD periods, shown in
Figure~\ref{f:fig3}.
These confirm that the hardening process does not preferentially act on more
massive binaries. 

\begin{figure}
\epsscale{.8}
\plotone{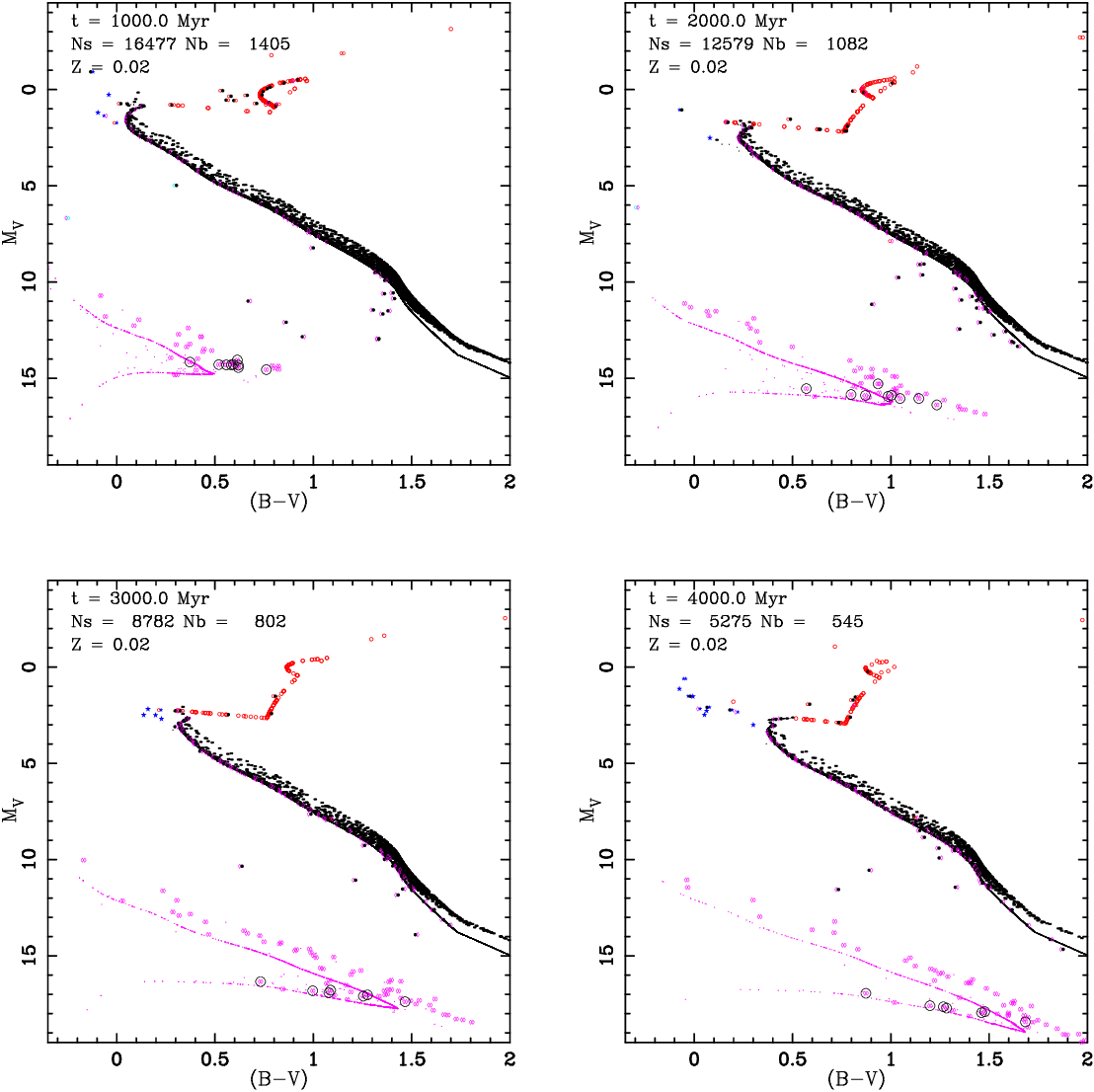}
\caption{
Colour-magnitude diagrams for the $Z = 0.004$ simulations at $1.0$, 
$2.0$, $3.0$ and $4.0\,$Gyr.
Stars from both simulations are plotted but the numbers of single stars 
and binaries given in each figure are averaged. 
Main-sequence stars (dots), blue stragglers (stars),
sub-giants, giants and naked helium stars (open circles)
and white dwarfs (dots) are distinguished.
Binary stars are denoted by overlapping symbols appropriate
to the stellar type of the components, with main-sequence binary components
depicted with filled circles and white dwarf binary components as diamonds.
Type Ia candidates are circled.
Bolometric corrections computed by \citet{kur92} from synthetic stellar
spectra are used to convert theoretical stellar quantities to observed
colours.
These corrections are strictly not valid for WDs and extremely cool giants.
\label{f:fig4}
}
\end{figure}

\begin{figure}
\epsscale{1.0}
\plotone{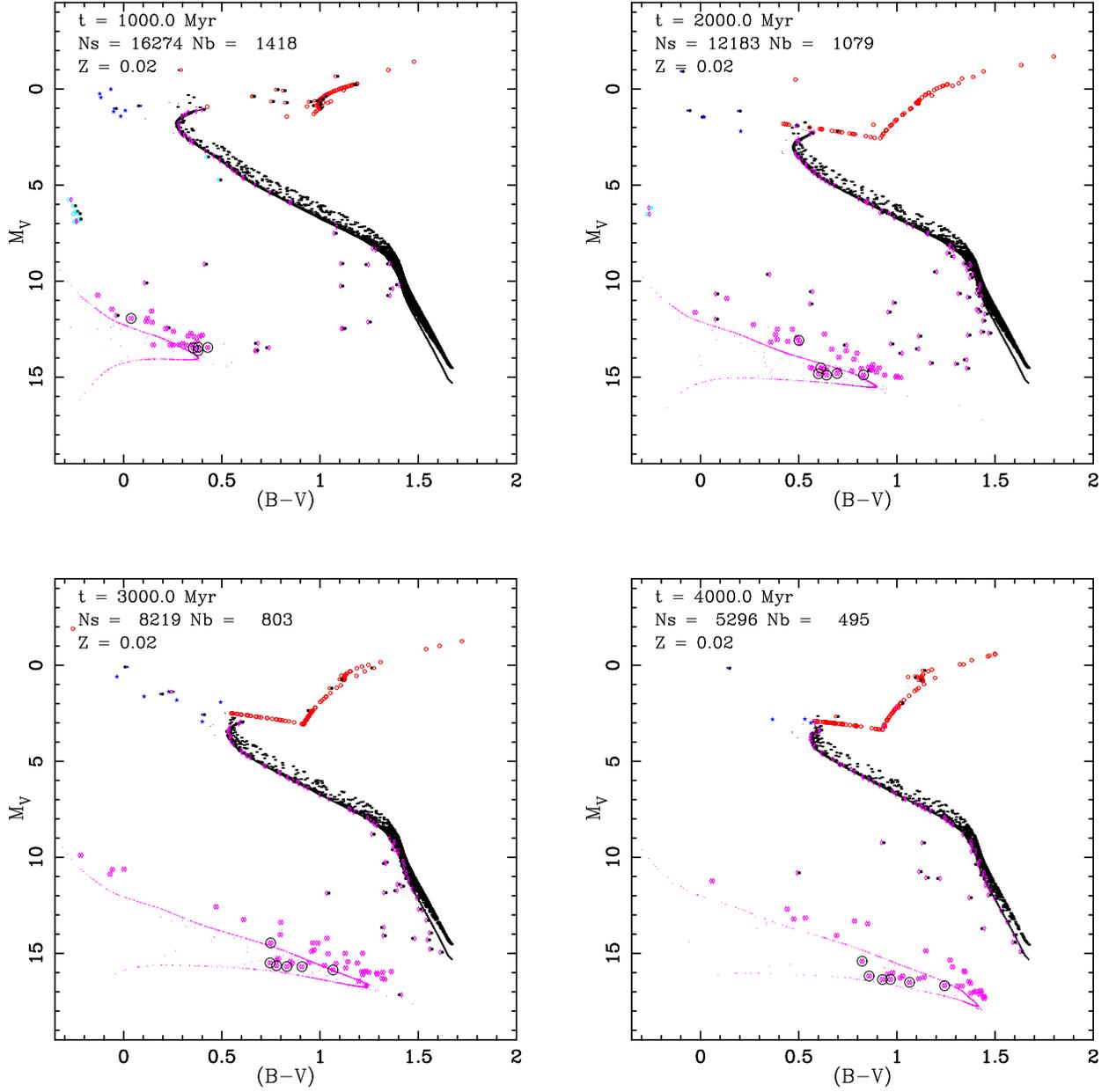}
\caption{
Same as Figure~\ref{f:fig4} but for the $Z = 0.02$ simulations. 
Note that many of the single main-sequence stars are overlayed by 
binary stars (in Figure~\ref{f:fig4} as well). 
\label{f:fig5}
}
\end{figure}

The color-magnitude diagrams of the clusters we have simulated are shown in 
Figures~\ref{f:fig4}-\ref{f:fig5} for the $Z = 0.004$ and $Z = 0.02$ models 
at $1\,$Gyr intervals. 
In addition to the single star and binary main-sequences, several blue stragglers 
and cataclysmic variables (CVs, below and to the left of the MS) are visible 
in each simulation. 
The latter have hardly been searched for in open clusters, 
and this work suggests that systematic searches might be rewarded. 
Most remarkable are the DWDs, seen in profusion above the white dwarf cooling 
sequence, and the ``loaded guns'' shown as circled diamonds in each of the figures.

\begin{figure}
\plotone{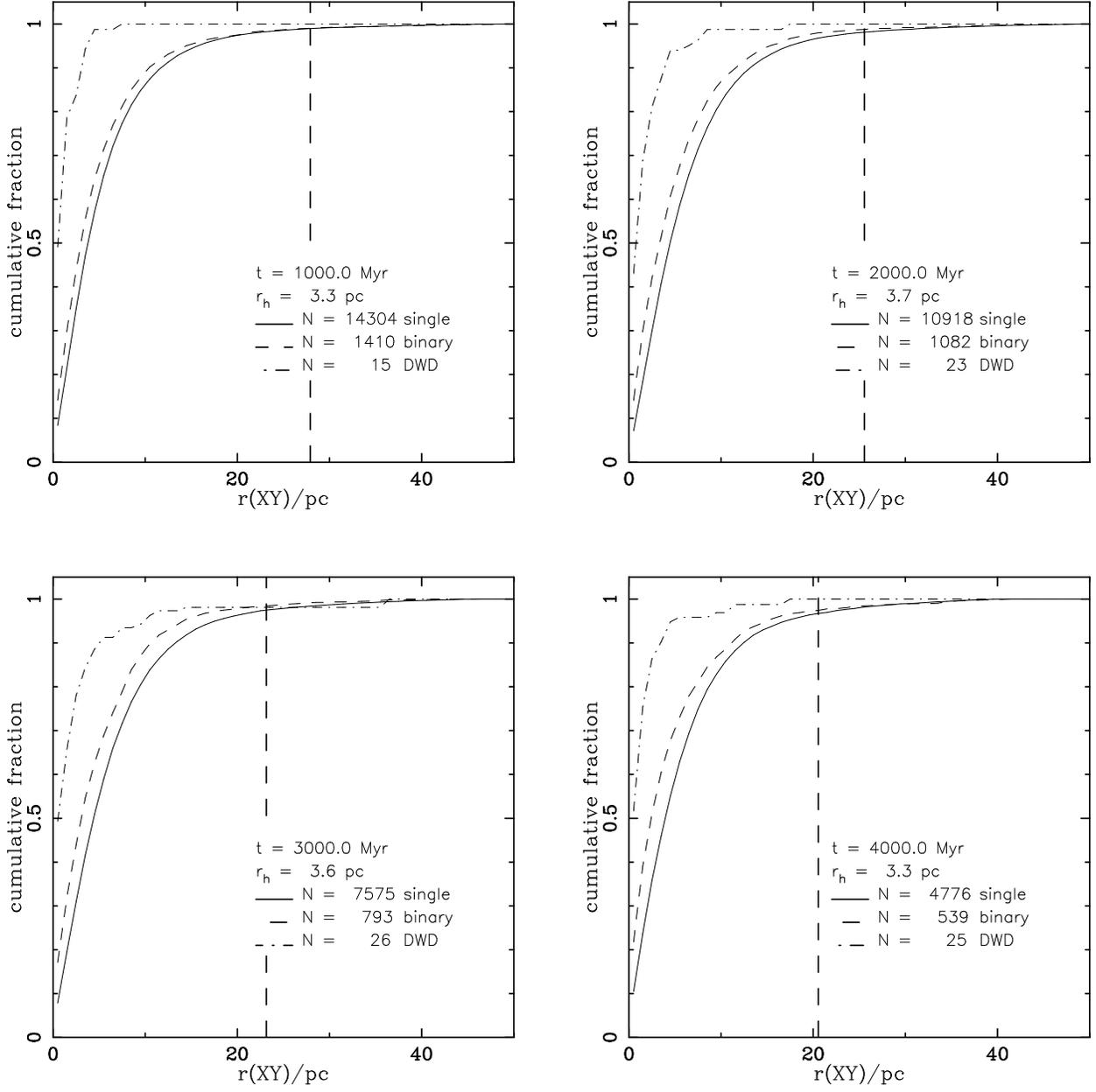}
\caption{
Population gradients in the XY-plane for single stars, binaries and double-WDs, 
averaged over the four simulations presented in this paper. 
The tidal radius of the cluster is shown as a vertical dashed line. 
Note that stars are not actually removed from a simulation until they 
are at a distance greater than two tidal radii from the cluster centre. 
The half-mass radius, $r_{\rm h}$, the tidal radius, 
and the numbers of each sub-population, 
are averaged values per simulation. 
\label{f:fig6}
}
\end{figure}

Figure~\ref{f:fig6} illustrates the spatial distribution within the cluster 
at several epochs for the single stars, binaries and DWDs. 
This clearly demonstrates that the quest for equipartition of energy is dominating 
the dynamical evolution: heavy objects segregate towards the inner regions of 
the cluster while lighter objects move outwards. 
Binaries are on average more massive than single stars and are therefore 
more centrally concentrated. 
Furthermore, the progenitor of any DWD was originally at least twice as massive 
as the cluster MS turn-off mass at the time the DWD formed, i.e. the DWDs 
form from the high-end of the binary IMF. 
This explains why DWDs are more centrally concentrated than the other binaries 
and means that they are most likely to be found in the core of the cluster, 
especially those with $M_{\rm b} > M_{\rm Ch}$.

\section{Accreting Single White Dwarfs}
\label{s:result2}

Table~\ref{t:table3} lists the possible ASDs generated in our simulations.
As with the DWDs, we see that about 2/3 of these
binary systems have been involved in exchange interactions. 
Most are giant-CO-WD pairs, but examples of He and ONe
white dwarfs orbiting giants are also present. 
The number of ASDs is comparable to that predicted by population synthesis 
without dynamical interactions ($\sim 6$ per simulation) but considering that 
only 1/3 of the ASDs evolved from primordial binaries it appears that 
dynamical interactions are destroying as many systems as are being created.

The ASDs are only listed as {\it possible} because they all have $q > 1$ 
which, according to most prescriptions for mass transfer, will lead to 
common-envelope evolution and not steady transfer of material onto the WD 
(see Section~\ref{s:method}). 
For these systems to become super-soft sources (and possibly SNIa) 
the common-envelope phase must be avoided, 
otherwise a DWD, or possibly a single giant, will result. 
The number of possible ASDs with $q < 1$ at the onset of mass transfer 
predicted by population synthesis is only $\sim 0.4$ per simulation which 
agrees with what we found. 
However, there is a fair amount of uncertainty in the value of $q_{\rm crit}$, 
the mass-ratio above which mass transfer from a giant proceeds on a dynamical 
timescale.  
\citet{web88} presents an expression for $q_{\rm crit}$ which differs by more 
than a factor of two compared to the model of \citet{hur02} for certain values 
of the giant envelope mass, and which gives $q_{\rm crit} > 2$ as the envelope 
mass becomes small. 
Furthermore, \citet{yun98} have shown that the assumption of a strong optically 
thick wind from the accreting star can act to stabilize the mass transfer. 
If we allow all values of $q_{\rm crit}$ to lead to stable mass transfer 
(on a thermal timescale), and assume that all of the transferred material is 
accreted by the WD, then Table~3 shows that $\sim 8$ WDs per simulation 
will reach the $M_{\rm Ch}$ and explode.

\section{Discussion}

The key result of this paper is that each of the possible progenitors of SNIa 
is preferentially manufactured in star clusters by orbital hardening and/or 
exchange interactions.
Consequently, in the dense environment of a star cluster, it is binaries altered by, 
or created from, dynamical interactions that provide the dominant formation channel 
for short-period DWDs and possible ASDs. 

\subsection{SNIa Locations} 

An overly simple prediction based on our results is that we might thus expect 
to see many or most Type Ia supernovae erupting in star clusters, 
often in their cores. 
There are two reasons to be cautious with this prediction.

First, open star clusters disperse, usually on a $1 - 6\,$Gyr timescale. 
Stars evaporate from clusters as they acquire 
energy from other cluster stars, as the Galactic tidal field 
incessantly tugs at cluster outliers, and through encounters with 
Giant Molecular Clouds. 
Second, we don't know what fraction of stars are created in star clusters, 
and what fraction of those escape during the earliest stages of cluster life. 
Observationally checking this prediction is difficult because only 
a dozen or so SNeIa have been identified closer than the Coma Cluster. 
Crowding will make at least some SNeIa appear ``close''
to star clusters, even if they are separated by hundreds of parsecs.

\subsection{Globular Clusters} 

At first glance, performing simulations of globular clusters should act to 
amplify the effects of dynamical interactions on the stellar populations 
within the cluster. 
These simulations will operate at higher particle density and thus the 
rate of stellar encounters will increase. 
However, in some cases, this may have the effect of closing 
production channels. 
Consider the ASD systems which comprise a Roche-lobe filling giant star and a 
WD secondary. 
For such a system to evolve to become a SNIa mass transfer onto the WD must 
be stable for a significant length of time. 
Observations of globular clusters \citep[for example]{guh98} show that 
bright giants are depleted in the core of the cluster indicating that 
they have been involved in collisions, a consequence of their relatively 
large cross-section. 
Such collisions could act to reduce the incidence of stable mass transfer from 
giants in a globular cluster, either by interrupting the mass transfer or 
preventing it from occurring at all. 
On the other hand, short-period DWD systems present a relatively small 
cross-section for collision so the increased stellar density should lead 
to an increased number of weak encounters. 
The orbital perturbations resulting from these will enhance the likelihood of 
DWD merger events. 

\subsection{SNIa Birthrates} 

The predicted Galactic birthrate of SNeIa from merging supra-Chandrasekhar DWDs 
is found by \citet{hur02} to be $2.6 \times 10^{-3} \, {\rm yr}^{-1}$. 
This number is calculated from population synthesis of isolated binaries 
using the same input parameters to the binary evolution model as used in 
our {\tt NBODY4} simulations. 
Alternatively, if the Gaussian period distribution for nearby solar-like stars 
found by \citet{duq91} is used to determine the initial orbits of the binaries, 
this number changes by less than 4\%. 
A similar rate is found by \citet{tut94} using an independent binary evolution 
algorithm. 
The observed rate is $4 \pm 1 \times 10^{-3} \, {\rm yr}^{-1}$ \citep{cap97} 
so clearly a factor of 10 increase in the predicted rate would bring 
binary evolution models into conflict with observations. 
However, \citet{hur02} found that if they draw the binary component masses 
independently from a single star IMF, rather than using a uniform mass-ratio 
distribution, the predicted rate drops by at least a factor of 10. 
So the enhancement of SNIa candidates found in our open cluster simulations 
would allow agreement for the more general mass-ratio distribution 
when trying to match the results of binary population synthesis studies 
to observations. 

There are additional parameters intrinsic to the binary evolution
algorithm which can also affect the production rate of DWDs.
The common-envelope efficiency parameter is a prime example.
Reducing ${\alpha}_{\rm CE}$ from 3.0 to 1.0 makes it harder to form
short-period binaries via the common-envelope channel and indeed,
\citet{hur02} found that this also decreased the predicted birthrate
of SNeIa from merging supra-Chandrasekhar DWDs by a factor of 10.
Because the critical mass-ratio for dynamical mass transfer affects the
frequency of common-envelope events, uncertainty in its value (see previous section)
will similarly affect any predicted number.
Trying to constrain the input parameters to a model of
binary evolution by matching the results of binary population synthesis to
observations of particular stellar populations is a risky business.
The number of uncertain parameters is large and it is entirely possible that an
error in one value will mask the error in another.
But the key result of this paper remains: hardening of binaries occurs only in 
clusters, and this preferentially creates SNIa candidates. 

\subsection{Binary Fraction} 

Clearly the number of DWD binaries produced in the simulations would increase
if we had used a higher binary fraction.
The fraction of 10\% is a rather conservative choice and we note that in some
open clusters it has been found that binaries may be as, or even more, populous
than single stars (Fan et al. 1996, for example).
We cannot stress enough however (this goes for the discussion in the previous
paragraph as well) that it is the number of close DWDs produced in
the simulations {\it relative} to the number produced in the field that is of
primary interest in this work.
Binaries do present a greater cross-section for dynamical
interaction than do single stars \citep{heg96} so a higher binary fraction does
have the capacity to affect the simulation results.
This will be addressed in future work.

\subsection{Eccentricity} 

The treatment of gravitational radiation used in the binary evolution algorithm
contains a strong dependence on orbital eccentricity, i.e. the removal of
angular momentum from the short-period system is accelerated if the orbit is
eccentric.
Binaries emerging from a common-envelope phase are assumed to be circular so the
only way for a short-period DWD orbit to be eccentric is through dynamical
interactions subsequent to DWD formation.
However, inducing an eccentricity in to an already circular orbit, especially in
the case of small orbital separation, is difficult \citep{hnr96}.
In none of the four simulated SNIa candidates that required perturbations to the
orbit of a DWD was a detectable eccentricity induced as
part of the hardening process.

On a related matter, \citet{hur01} have shown that the shape of the eccentricity
distribution used for the primordial binary population can affect the number of
blue stragglers produced from that population, for example.
This is less important in the case of DWD binaries as the systems of 
interest will simply form from a slightly different set of primordial binaries. 
The same goes for variations in the way that tidal friction is modelled
(see Hurley, Tout \& Pols 2002 for a detailed discussion).

\subsection{DWDs and CVs in Clusters} 

Another intriguing suggestion from this work is that DWDs of all periods should 
be rather commonplace in open clusters, and likely segregated towards the 
cluster centres. 
A significant fraction of these DWDs must have been involved in exchange 
interactions. 
We encourage observers to survey for such objects. 
In the case of CVs we do not find a similar 
enhancement in open clusters: typically $1-2$ CVs were formed in each 
simulation. 
We expect that this is mainly due to the required presence of a low-mass 
MS star, less massive than its WD companion, in any binary that will evolve 
to a long-lived, and thus stable, CV state. 
This means that the progenitors of CVs will on average be less massive than the 
progenitors of DWDs, either on the ZAMS or after the WD (or WDs) have formed, 
so that they are less likely to reside in the core of the cluster and be 
involved in dynamical interactions. 
Furthermore, in a 3-body exchange interaction, it is normally the least massive 
star that is ejected so CVs are unlikely to form in this manner. 
We do not rule out the possiblity that CV production will be enhanced in 
simulations of globular clusters. 
We also note that if merging supra-Chandrasekhar DWDs lead to the formation of 
neutron stars via an AIC then they are still extremely interesting objects in 
terms of the neutron star retention problem in star clusters \citep{pfa02}. 

\section{Conclusions}
\label{s:conclu}

We find a greatly increased rate of production of ``loaded guns'' -
supra-Chandrasekhar double-white-dwarfs with inspiral ages shorter than a 
Hubble time - in star clusters relative to the field. 
Orbital hardening and exchange interactions are the responsible 
mechanisms for the enhanced rates. 
Neither of these processes operates in the field, and neither has been included 
in previous population studies of SNIa progenitors. 

The production rate of possible accreting single degenerates is not 
significantly enhanced relative to the field. 
However, a major fraction of these systems are formed in exchange interactions, 
which means that modelling of the stellar dynamics is destroying production 
channels as well as creating them. 
Whether or not accreting single degenerates will evolve to become 
super-soft sources, and ultimately type Ia supernovae, depends critically 
on avoiding the phase of common-envelope evolution that occurs if 
mass transfer proceeds on a dynamical timescale. 

The results are based on studies of open cluster size $N$-body simulations, 
but we expect the effect to be even stronger in the case of globular clusters. 
The more frequent encounters in globulars will harden DWDs much more 
rapidly than in open clusters, and hence increase the predicted 
rate of creation of SNIa progenitors.
An increased rate of collisions involving giant stars will deplete the numbers 
of accreting single degenerate binaries.

\acknowledgments

We are extremely grateful to Jun Makino and the University of Tokyo for the 
loan of the GRAPE-6 board. 
The generous support of Doug Ellis and the Cordelia Corporation 
has enabled AMNH to purchase new GRAPE-6 boards and we are most 
grateful for that. 
MS thanks Ken Nomoto for very helpful discussions and suggestions. 
We thank Fred Rasio and Vicky Kalogera for organizing the 2001 Aspen Center 
for Physics meeting on Star Clusters where some of the ideas for this paper 
were clarified.
We also thank the referee for comments that greatly improved certain
aspects of this manuscript.

\clearpage

\begin{deluxetable}{rllrrrlll}
\tablecolumns{9}
\tablewidth{0pc}
\tablecaption{
Double-WD systems that are Type Ia candidates.
To qualify the system must have a combined mass in excess of the
Chandrasekhar mass, $1.44 M_{\odot}$, and a gravitational
radiation merger timescale less than the age of the Universe,
$\sim 1.2 \times 10^{10} \,$yr.
The time at formation for the double-WD (DWD) system, Myr units, is given in Column~1.
The types of the WDs are listed in Column~2.
Three types of WD are distinguished: helium composition (He),
carbon-oxygen (CO), and oxygen-neon (ONe).
The individual masses of the two WDs are given in Columns~3 and 4
respectively, and the combined mass is given in Column~5.
All masses are in solar units.
The period of the binary is given in Column~6 in units of days.
Column~7 gives an estimate of the time it will take the DWD system
to merge, in yrs, owing to angular momentum loss from
gravitational radiation.
This estimate comes from integrating eq.~(48) of Hurley, Tout \& Pols (2002).
Column~8 summarizes the history of each system using the following code:
primordial binary (PRIM); exchange interaction (EXCH);
perturbation before (PB), or after (PA), Double-WD formed;
subsequent escape from cluster (ESC); subsequent disruption in
dynamical encounter (DISR).
Note that perturbations to the orbit are only recorded if they
lead to a change in the evolution path of the binary.
\label{t:table1}
}
\tabletypesize\footnotesize
\tablehead{
$T_{\rm form}$ & \multicolumn{2}{c}{Types} & $M_1$ & $M_2$ & $M_{\rm b}$ &
\multicolumn{1}{c}{Period} & \multicolumn{1}{c}{$T_{\rm grav}$} & Legend
}
\startdata
 225 & CO  & ONe & 0.72 & 1.24 & 1.96 & $3.3884 \times 10^{-1}$ & $4.112 \times 10^{9}$ & EXCH \\
 186 & CO  & CO & 0.99 & 0.66 & 1.65 & $1.0715 \times 10^{-1}$ & $2.259 \times 10^{8}$ & PRIM-PB-ESC \\
 229 & CO  & CO & 0.97 & 0.67 & 1.64 & $1.1482 \times 10^{-1}$ & $2.991 \times 10^{8}$ & PRIM-PA-DISR \\
 334 & ONe & CO & 1.06 & 0.57 & 1.63 & $1.0715 \times 10^{-1}$ & $2.270 \times 10^{8}$ & PRIM-PA \\
 370 & CO  & CO & 1.09 & 0.54 & 1.63 & $2.4547 \times 10^{-1}$ & $2.343 \times 10^{9}$ & PRIM-PA \\
 221 & CO  & CO & 0.92 & 0.64 & 1.56 & $4.3652 \times 10^{-2}$ & $2.187 \times 10^{7}$ & PRIM-ESC \\
 375 & CO  & CO & 0.83 & 0.67 & 1.50 & $1.1220 \times 10^{-1}$ & $2.997 \times 10^{8}$ & PRIM \\
 149 & CO  & CO & 0.83 & 0.66 & 1.49 & $8.7096 \times 10^{-3}$ & $3.350 \times 10^{5}$ & PRIM-PB-ESC \\
\tableline
 223 & CO  & CO & 0.97 & 0.73 & 1.70 & $4.8722 \times 10^{-1}$ & $1.195 \times 10^{10}$ & PRIM \\
1299 & CO  & CO & 1.07 & 0.46 & 1.53 & $5.1286 \times 10^{-2}$ & $4.563 \times 10^{7}$ & EXCH \\
\tableline
 123 & CO  & CO & 1.16 & 0.68 & 1.84 & $1.0965 \times 10^{-2}$ & $4.633 \times 10^{5}$ & PRIM-ESC \\
 112 & ONe & CO & 1.10 & 0.67 & 1.77 & $1.6596 \times 10^{-2}$ & $1.544 \times 10^{6}$ & PRIM-DISR \\
 225 & CO  & CO & 0.95 & 0.73 & 1.68 & $4.7839 \times 10^{-1}$ & $1.169 \times 10^{10}$ & PRIM-PA \\
\tableline
 223 & CO  & CO & 1.09 & 0.71 & 1.80 & $4.6800 \times 10^{0}$ & $1.007 \times 10^{10}$ & PRIM-PB \\
 259 & ONe & CO & 1.12 & 0.66 & 1.78 & $4.0738 \times 10^{-1}$ & $6.958 \times 10^{9}$ & PRIM \\
1192 & ONe & CO & 1.29 & 0.30 & 1.59 & $3.4449 \times 10^{-1}$ & $1.148 \times 10^{10}$ & EXCH \\
\enddata
\end{deluxetable}

\clearpage

\begin{deluxetable}{rllrrrlll}
\tablecolumns{9}
\tablewidth{0pc}
\tablecaption{
Double-WD systems that are not Type Ia candidates.
Formed either from a primordial binary (PRIM) or via an
exchange (EXCH) interaction.
See Table~\ref{t:table1} for an explanation of what each column entails. 
\label{t:table2}
}
\tabletypesize\footnotesize
\tablehead{
$T_{\rm form}$ & \multicolumn{2}{c}{Types} & $M_1$ & $M_2$ & $M_{\rm b}$ &
\multicolumn{1}{c}{Period} & \multicolumn{1}{c}{$T_{\rm grav}$} & Legend
}
\startdata
\cutinhead{Simulation~1: $Z = 0.004$}
 334 & ONe & CO & 1.30 & 0.88 & 2.18 & $2.2909 \times 10^{4}$ & $6.804 \times 10^{23}$ & PRIM \\
 260 & CO  & CO & 1.08 & 0.89 & 1.97 & $6.3096 \times 10^{3}$ & $6.575 \times 10^{21}$ & PRIM \\
4358 & CO  & CO & 0.62 & 1.26 & 1.88 & $3.6308 \times 10^{4}$ & $8.779 \times 10^{23}$ & EXCH \\
2227 & CO  & CO & 0.89 & 0.90 & 1.79 & $3.4674 \times 10^{5}$ & $3.417 \times 10^{26}$ & EXCH \\
 557 & CO  & CO & 0.97 & 0.79 & 1.76 & $2.2910 \times 10^{2}$ & $1.160 \times 10^{18}$ & EXCH \\
2858 & CO  & CO & 0.70 & 1.01 & 1.71 & $3.0903 \times 10^{5}$ & $2.668 \times 10^{26}$ & EXCH \\
3600 & CO  & CO & 0.97 & 0.72 & 1.69 & $1.8621 \times 10^{2}$ & $7.233 \times 10^{17}$ & EXCH \\
1819 & ONe & He & 1.29 & 0.37 & 1.66 & $5.4954 \times 10^{0}$ & $8.884 \times 10^{13}$ & EXCH \\
2116 & CO  & CO & 1.06 & 0.59 & 1.65 & $1.7378 \times 10^{2}$ & $6.759 \times 10^{17}$ & EXCH \\
1001 & CO  & CO & 0.71 & 0.87 & 1.58 & $8.9125 \times 10^{3}$ & $2.465 \times 10^{22}$ & EXCH \\
 596 & CO  & CO & 0.69 & 0.82 & 1.51 & $5.1286 \times 10^{0}$ & $5.872 \times 10^{13}$ & EXCH \\
1596 & CO  & CO & 0.83 & 0.66 & 1.49 & $1.0001 \times 10^{6}$ & $2.352 \times 10^{28}$ & EXCH \\
4232 & CO  & CO & 1.07 & 0.41 & 1.48 & $1.9953 \times 10^{1}$ & $2.944 \times 10^{15}$ & EXCH \\
 594 & CO  & CO & 0.79 & 0.67 & 1.46 & $5.7544 \times 10^{-1}$ & $2.157 \times 10^{10}$ & EXCH \\
1522 & CO  & CO & 0.71 & 0.71 & 1.42 & $8.3176 \times 10^{4}$ & $1.102 \times 10^{25}$ & EXCH \\
4121 & CO  & CO & 0.67 & 0.63 & 1.30 & $5.6234 \times 10^{5}$ & $2.079 \times 10^{27}$ & EXCH \\
1530 & CO  & CO & 0.59 & 0.70 & 1.29 & $2.1380 \times 10^{-1}$ & $2.209 \times 10^{9}$ & EXCH \\
2933 & CO  & CO & 0.70 & 0.59 & 1.29 & $2.5119 \times 10^{2}$ & $2.515 \times 10^{18}$ & PRIM \\
 786 & CO  & CO & 0.67 & 0.61 & 1.28 & $1.7378 \times 10^{-1}$ & $1.098 \times 10^{9}$ & EXCH \\
1009 & CO  & CO & 0.73 & 0.47 & 1.20 & $1.4791 \times 10^{-1}$ & $9.000 \times 10^{8}$ & EXCH \\
2784 & CO  & CO & 0.54 & 0.61 & 1.15 & $5.3703 \times 10^{1}$ & $4.963 \times 10^{16}$ & EXCH \\
2487 & CO  & CO & 0.69 & 0.43 & 1.12 & $1.0715 \times 10^{1}$ & $7.380 \times 10^{14}$ & PRIM \\
 631 & He  & CO & 0.34 & 0.77 & 1.11 & $2.5704 \times 10^{-3}$ & $2.507 \times 10^{4}$ & EXCH \\
2494 & CO  & CO & 0.47 & 0.63 & 1.10 & $2.0417 \times 10^{1}$ & $4.123 \times 10^{15}$ & EXCH \\
1930 & CO  & He & 0.68 & 0.40 & 1.08 & $1.6596 \times 10^{0}$ & $8.215 \times 10^{11}$ & EXCH \\
1499 & CO  & He & 0.67 & 0.33 & 1.00 & $1.2023 \times 10^{-1}$ & $8.265 \times 10^{8}$ & EXCH \\
2079 & CO  & CO & 0.35 & 0.64 & 0.99 & $2.0893 \times 10^{-2}$ & $7.279 \times 10^{6}$ & PRIM \\
1690 & He  & CO & 0.36 & 0.62 & 0.98 & $1.9055 \times 10^{0}$ & $9.372 \times 10^{12}$ & EXCH \\
4343 & CO  & CO & 0.77 & 0.20 & 0.97 & $1.2303 \times 10^{-1}$ & $1.223 \times 10^{9}$ & PRIM \\
 601 & He  & CO & 0.04 & 0.92 & 0.96 & $1.4791 \times 10^{-2}$ & $1.955 \times 10^{7}$ & PRIM \\
3898 & CO  & CO & 0.40 & 0.55 & 0.95 & $7.4131 \times 10^{0}$ & $3.597 \times 10^{14}$ & EXCH \\
1671 & He  & CO & 0.32 & 0.62 & 0.94 & $8.5114 \times 10^{-1}$ & $1.450 \times 10^{11}$ & PRIM \\
3007 & CO  & CO & 0.66 & 0.27 & 0.93 & $1.9055 \times 10^{-1}$ & $4.164 \times 10^{9}$ & EXCH \\
2153 & CO  & CO & 0.55 & 0.35 & 0.90 & $1.9498 \times 10^{0}$ & $1.108 \times 10^{13}$ & EXCH \\
2636 & CO  & CO & 0.60 & 0.30 & 0.90 & $8.1283 \times 10^{-1}$ & $1.597 \times 10^{11}$ & EXCH \\
 780 & He  & CO & 0.33 & 0.51 & 0.84 & $1.5849 \times 10^{-2}$ & $4.211 \times 10^{6}$ & PRIM \\
3267 & CO  & CO & 0.68 & 0.30 & 0.98 & $1.1749 \times 10^{0}$ & $3.638 \times 10^{11}$ & EXCH \\
3341 & CO  & CO & 0.32 & 0.54 & 0.86 & $1.5849 \times 10^{0}$ & $8.390 \times 10^{11}$ & EXCH \\
3564 & He  & CO & 0.20 & 0.53 & 0.73 & $7.5858 \times 10^{-2}$ & $3.774 \times 10^{8}$ & EXCH \\
3601 & CO  & CO & 0.48 & 0.19 & 0.67 & $3.9811 \times 10^{-2}$ & $8.338 \times 10^{7}$ & EXCH \\
\cutinhead{Simulation~2: $Z = 0.004$}
 111 & CO  & CO & 1.21 & 1.14 & 2.35 & $1.5849 \times 10^{4}$ & $5.843 \times 10^{22}$ & PRIM \\
 408 & CO  & CO & 1.08 & 0.87 & 1.95 & $2.5119 \times 10^{4}$ & $2.769 \times 10^{23}$ & PRIM \\
 259 & CO  & CO & 1.01 & 0.74 & 1.75 & $1.7783 \times 10^{0}$ & $3.461 \times 10^{11}$ & PRIM \\
3489 & CO  & CO & 0.64 & 1.08 & 1.72 & $3.1623 \times 10^{4}$ & $6.341 \times 10^{23}$ & EXCH \\
 668 & CO  & CO & 0.77 & 0.89 & 1.67 & $1.3183 \times 10^{4}$ & $6.171 \times 10^{22}$ & PRIM \\
 259 & CO  & CO & 0.93 & 0.73 & 1.66 & $1.5136 \times 10^{0}$ & $2.366 \times 10^{11}$ & PRIM \\
2449 & CO  & CO & 0.86 & 0.78 & 1.64 & $3.9811 \times 10^{1}$ & $1.189 \times 10^{16}$ & EXCH \\
 593 & CO  & CO & 0.75 & 0.88 & 1.62 & $6.3096 \times 10^{1}$ & $4.279 \times 10^{16}$ & PRIM \\
2672 & CO  & CO & 0.61 & 0.93 & 1.55 & $3.9811 \times 10^{4}$ & $1.368 \times 10^{24}$ & EXCH \\
 890 & CO  & CO & 0.88 & 0.66 & 1.54 & $6.6069 \times 10^{1}$ & $5.497 \times 10^{16}$ & PRIM \\
 482 & ONe & CO & 1.04 & 0.49 & 1.53 & $9.7724 \times 10^{-1}$ & $1.152 \times 10^{11}$ & PRIM \\
 668 & CO  & CO & 0.86 & 0.65 & 1.52 & $1.2882 \times 10^{1}$ & $7.099 \times 10^{14}$ & EXCH \\
1187 & CO  & CO & 0.69 & 0.80 & 1.49 & $6.1659 \times 10^{2}$ & $2.105 \times 10^{19}$ & EXCH \\
1410 & CO  & CO & 0.78 & 0.68 & 1.46 & $1.8197 \times 10^{4}$ & $1.819 \times 10^{23}$ & EXCH \\
1373 & CO  & CO & 0.57 & 0.86 & 1.43 & $3.0200 \times 10^{1}$ & $7.814 \times 10^{15}$ & EXCH \\
2078 & CO  & CO & 0.78 & 0.63 & 1.41 & $1.5849 \times 10^{4}$ & $1.394 \times 10^{23}$ & EXCH \\
1781 & CO  & CO & 0.75 & 0.65 & 1.40 & $1.9498 \times 10^{2}$ & $1.107 \times 10^{18}$ & EXCH \\
 222 & CO  & CO & 0.64 & 0.70 & 1.34 & $6.1659 \times 10^{-2}$ & $7.386 \times 10^{7}$ & PRIM \\
2821 & CO  & CO & 0.71 & 0.62 & 1.34 & $1.5136 \times 10^{4}$ & $1.289 \times 10^{23}$ & EXCH \\
 259 & CO  & CO & 0.60 & 0.72 & 1.32 & $1.3804 \times 10^{-1}$ & $6.282 \times 10^{8}$ & PRIM \\
1187 & CO  & CO & 0.74 & 0.58 & 1.32 & $4.0738 \times 10^{0}$ & $3.938 \times 10^{13}$ & EXCH \\
2301 & CO  & CO & 0.64 & 0.62 & 1.26 & $2.6303 \times 10^{4}$ & $6.209 \times 10^{23}$ & EXCH \\
1893 & ONe & He & 1.01 & 0.24 & 1.25 & $3.0200 \times 10^{-1}$ & $8.483 \times 10^{9}$ & PRIM \\
2412 & CO  & CO & 0.62 & 0.63 & 1.25 & $9.7724 \times 10^{4}$ & $2.049 \times 10^{25}$ & EXCH \\
1336 & CO  & He & 0.88 & 0.37 & 1.24 & $2.6303 \times 10^{0}$ & $1.626 \times 10^{13}$ & PRIM \\
2115 & CO  & CO & 0.68 & 0.54 & 1.22 & $2.4547 \times 10^{1}$ & $5.731 \times 10^{15}$ & EXCH \\
 408 & CO  & CO & 0.57 & 0.63 & 1.21 & $9.5499 \times 10^{-2}$ & $2.886 \times 10^{8}$ & PRIM \\
1410 & CO  & CO & 0.53 & 0.68 & 1.21 & $1.4125 \times 10^{-1}$ & $7.178 \times 10^{8}$ & EXCH \\
 482 & CO  & CO & 0.61 & 0.57 & 1.18 & $1.9055 \times 10^{-1}$ & $1.640 \times 10^{9}$ & PRIM \\
1707 & CO  & CO & 0.83 & 0.34 & 1.18 & $8.3176 \times 10^{-1}$ & $1.010 \times 10^{11}$ & PRIM \\
2524 & CO  & CO & 0.48 & 0.65 & 1.13 & $1.0000 \times 10^{0}$ & $1.798 \times 10^{11}$ & EXCH \\
2190 & CO  & He & 0.72 & 0.32 & 1.04 & $1.2882 \times 10^{0}$ & $4.787 \times 10^{11}$ & EXCH \\
 853 & CO  & CO & 0.40 & 0.62 & 1.03 & $4.3652 \times 10^{-1}$ & $2.404 \times 10^{10}$ & PRIM \\
1967 & CO  & He & 0.71 & 0.31 & 1.02 & $2.7542 \times 10^{-1}$ & $7.620 \times 10^{9}$ & EXCH \\
 965 & He  & CO & 0.27 & 0.73 & 1.00 & $1.1220 \times 10^{-1}$ & $7.299 \times 10^{8}$ & EXCH \\
3489 & CO  & He & 0.58 & 0.41 & 0.99 & $8.9125 \times 10^{0}$ & $5.281 \times 10^{14}$ & PRIM \\
1893 & He  & CO & 0.34 & 0.58 & 0.92 & $1.5488 \times 10^{0}$ & $6.997 \times 10^{11}$ & EXCH \\
3971 & He  & He & 0.56 & 0.34 & 0.90 & $3.0903 \times 10^{-3}$ & $5.145 \times 10^{4}$ & EXCH \\
2338 & CO  & He & 0.57 & 0.31 & 0.88 & $8.1283 \times 10^{-1}$ & $1.649 \times 10^{11}$ & EXCH \\
1744 & CO  & He & 0.56 & 0.31 & 0.87 & $7.2444 \times 10^{-1}$ & $1.193 \times 10^{11}$ & EXCH \\
1818 & CO  & He & 0.64 & 0.20 & 0.83 & $3.0200 \times 10^{-2}$ & $3.238 \times 10^{7}$ & EXCH \\
2709 & He  & He & 0.47 & 0.33 & 0.79 & $1.2023 \times 10^{0}$ & $5.395 \times 10^{11}$ & EXCH \\
4046 & He  & He & 0.38 & 0.23 & 0.61 & $1.3490 \times 10^{-1}$ & $2.522 \times 10^{9}$ & PRIM \\
2895 & He  & He & 0.32 & 0.20 & 0.52 & $3.4674 \times 10^{-2}$ & $7.083 \times 10^{7}$ & PRIM \\ 
\cutinhead{Simulation~3: $Z = 0.02$}
 112 & CO  & CO & 1.16 & 1.12 & 2.28 & $1.5488 \times 10^{3}$ & $1.242 \times 10^{20}$ & PRIM \\
 186 & CO  & CO & 1.01 & 0.90 & 1.91 & $5.0119 \times 10^{1}$ & $1.751 \times 10^{16}$ & PRIM \\
 261 & CO  & CO & 0.86 & 0.94 & 1.80 & $9.7724 \times 10^{3}$ & $2.419 \times 10^{22}$ & PRIM \\
 298 & CO  & CO & 0.59 & 0.63 & 1.21 & $1.4454 \times 10^{-1}$ & $8.038 \times 10^{8}$ & PRIM \\
 373 & CO  & CO & 0.79 & 0.81 & 1.60 & $1.4454 \times 10^{4}$ & $8.335 \times 10^{22}$ & EXCH \\
 560 & CO  & CO & 0.77 & 0.73 & 1.50 & $2.2387 \times 10^{4}$ & $2.994 \times 10^{23}$ & PRIM \\
 597 & CO  & CO & 0.62 & 0.71 & 1.33 & $1.3804 \times 10^{0}$ & $2.724 \times 10^{11}$ & EXCH \\
 597 & CO  & CO & 0.83 & 0.73 & 1.56 & $1.7783 \times 10^{4}$ & $1.577 \times 10^{23}$ & EXCH \\
 634 & CO  & CO & 0.83 & 0.62 & 1.45 & $6.1660 \times 10^{0}$ & $1.054 \times 10^{14}$ & PRIM \\
 672 & CO  & CO & 0.72 & 0.61 & 1.33 & $2.6915 \times 10^{0}$ & $1.338 \times 10^{13}$ & EXCH \\
 784 & CO  & CO & 0.70 & 0.69 & 1.39 & $1.6218 \times 10^{2}$ & $6.903 \times 10^{17}$ & EXCH \\
 896 & CO  & CO & 1.16 & 0.58 & 1.74 & $1.1482 \times 10^{1}$ & $4.486 \times 10^{14}$ & EXCH \\
 971 & CO  & CO & 0.75 & 0.65 & 1.41 & $7.4131 \times 10^{4}$ & $8.103 \times 10^{24}$ & EXCH \\
1008 & CO  & CO & 0.57 & 0.70 & 1.26 & $2.8184 \times 10^{0}$ & $1.631 \times 10^{13}$ & EXCH \\
1120 & CO  & CO & 0.55 & 0.67 & 1.23 & $9.1201 \times 10^{-1}$ & $1.095 \times 10^{11}$ & EXCH \\
1157 & CO  & He & 0.77 & 0.32 & 1.09 & $1.0000 \times 10^{-2}$ & $1.018 \times 10^{6}$ & EXCH \\
1195 & CO  & CO & 0.81 & 0.55 & 1.36 & $1.4125 \times 10^{1}$ & $1.098 \times 10^{15}$ & EXCH \\
1344 & CO  & CO & 0.83 & 0.83 & 1.67 & $2.0893 \times 10^{4}$ & $2.090 \times 10^{23}$ & EXCH \\
1643 & CO  & CO & 0.63 & 0.55 & 1.19 & $3.7154 \times 10^{1}$ & $1.697 \times 10^{16}$ & EXCH \\
1643 & CO  & CO & 0.62 & 0.63 & 1.25 & $1.7783 \times 10^{4}$ & $2.214 \times 10^{23}$ & EXCH \\
1643 & CO  & CO & 0.65 & 0.43 & 1.09 & $7.4131 \times 10^{-2}$ & $1.561 \times 10^{8}$ & EXCH \\
1680 & CO  & CO & 0.65 & 0.62 & 1.27 & $2.1380 \times 10^{4}$ & $3.548 \times 10^{23}$ & EXCH \\
2240 & CO  & CO & 0.89 & 0.34 & 1.23 & $9.5499 \times 10^{-1}$ & $1.652 \times 10^{11}$ & EXCH \\
2875 & He  & CO & 0.41 & 0.77 & 1.17 & $1.4125 \times 10^{1}$ & $1.466 \times 10^{15}$ & EXCH \\
2913 & He  & He & 0.28 & 0.41 & 0.69 & $4.4668 \times 10^{-1}$ & $3.984 \times 10^{10}$ & EXCH \\
3846 & He  & CO & 0.22 & 0.77 & 0.99 & $3.1623 \times 10^{-1}$ & $1.271 \times 10^{10}$ & EXCH \\
4108 & CO  & He & 0.71 & 0.32 & 1.03 & $2.6915 \times 10^{0}$ & $2.324 \times 10^{13}$ & PRIM \\
\cutinhead{Simulation~4: $Z = 0.02$}
 148 & CO  & CO & 1.14 & 1.00 & 2.14 & $6.6970 \times 10^{2}$ & $1.501 \times 10^{19}$ & PRIM \\
 445 & ONe & CO & 1.29 & 0.82 & 2.08 & $9.1018 \times 10^{3}$ & $1.671 \times 10^{22}$ & PRIM \\
 222 & CO  & CO & 1.12 & 0.89 & 2.01 & $1.6921 \times 10^{2}$ & $4.188 \times 10^{17}$ & PRIM \\
 222 & CO  & CO & 0.90 & 1.11 & 2.01 & $4.6799 \times 10^{1}$ & $1.360 \times 10^{16}$ & EXCH \\
 445 & CO  & CO & 1.10 & 0.84 & 1.94 & $3.6207 \times 10^{1}$ & $7.584 \times 10^{15}$ & PRIM \\
 743 & CO  & CO & 0.73 & 0.79 & 1.52 & $1.6410 \times 10^{4}$ & $1.335 \times 10^{23}$ & EXCH \\
 780 & CO  & CO & 0.70 & 0.71 & 1.41 & $4.3348 \times 10^{2}$ & $9.235 \times 10^{18}$ & EXCH \\
1449 & CO  & CO & 0.70 & 0.62 & 1.32 & $2.8185 \times 10^{4}$ & $7.184 \times 10^{23}$ & PRIM \\
 854 & He  & CO & 0.34 & 0.88 & 1.22 & $9.0059 \times 10^{-2}$ & $3.105 \times 10^{8}$ & EXCH \\
1189 & CO  & CO & 0.62 & 0.60 & 1.22 & $1.0313 \times 10^{-1}$ & $3.087 \times 10^{8}$ & EXCH \\
1151 & CO  & CO & 0.39 & 0.81 & 1.20 & $1.8924 \times 10^{-1}$ & $2.205 \times 10^{9}$ & EXCH \\
 706 & CO  & He & 0.89 & 0.30 & 1.19 & $8.9975 \times 10^{-2}$ & $3.105 \times 10^{8}$ & PRIM \\
1932 & CO  & CO & 0.58 & 0.61 & 1.19 & $5.6197 \times 10^{4}$ & $5.139 \times 10^{24}$ & EXCH \\
1709 & CO  & CO & 0.60 & 0.58 & 1.18 & $5.3005 \times 10^{0}$ & $9.472 \times 10^{13}$ & EXCH \\
 594 & CO  & CO & 0.56 & 0.59 & 1.15 & $1.0130 \times 10^{-1}$ & $3.087 \times 10^{8}$ & EXCH \\
1263 & CO  & He & 0.77 & 0.29 & 1.06 & $1.0082 \times 10^{0}$ & $2.086 \times 10^{11}$ & PRIM \\
3009 & He  & CO & 0.43 & 0.58 & 1.01 & $1.6502 \times 10^{1}$ & $2.762 \times 10^{15}$ & EXCH \\
1635 & He  & CO & 0.23 & 0.72 & 0.95 & $8.8876 \times 10^{-2}$ & $5.445 \times 10^{8}$ & EXCH \\
1523 & CO  & He & 0.59 & 0.33 & 0.92 & $1.6947 \times 10^{0}$ & $8.297 \times 10^{12}$ & EXCH \\
2229 & CO  & He & 0.60 & 0.31 & 0.91 & $5.1021 \times 10^{-1}$ & $4.095 \times 10^{10}$ & EXCH \\
2118 & CO  & CO & 0.29 & 0.59 & 0.88 & $6.0328 \times 10^{-1}$ & $6.660 \times 10^{10}$ & EXCH \\
2601 & CO  & He & 0.63 & 0.23 & 0.86 & $1.0000 \times 10^{-1}$ & $8.082 \times 10^{8}$ & EXCH \\
1820 & He  & He & 0.22 & 0.43 & 0.65 & $9.4180 \times 10^{-2}$ & $8.318 \times 10^{8}$ & EXCH \\
1151 & He  & He & 0.31 & 0.32 & 0.63 & $9.0011 \times 10^{-2}$ & $7.395 \times 10^{8}$ & PRIM \\
2935 & He  & He & 0.35 & 0.25 & 0.60 & $9.0417 \times 10^{-1}$ & $1.107 \times 10^{9}$ & EXCH \\
\enddata
\end{deluxetable}

\clearpage

\begin{deluxetable}{rllrrrrll}
\tablecolumns{9}
\tablewidth{0pc}
\tabletypesize{\small}
\tablecaption{
Potential super-soft sources.
Time at onset of mass transfer is given in Column~1, in Myr units, and
the stellar types of the stars at this time are listed in
Column~2: Roche-lobe filling star is either on the
Hertzsprung gap (HG) or the first giant branch (GB).
The mass of the sub-giant or giant, the mass of the WD, and the
mass-ratio ($q = M_1/M_2$), are given in Columns~3, 4, and 5,
respectively.
Column~6 shows the resultant mass of the WD if it were to accept all
of the mass available in the donor stars envelope at the onset of
mass-transfer.
All masses are in solar units.
The period of the binary is given in Column~7 in units of days.
Column~8 summarizes the history of each system using the following code:
primordial binary (PRIM); exchange interaction (EXCH);
perturbation to orbit before Roche-lobe overflow (PB);
steady mass-transfer (SMT); common-envelope evolution (CE).
The outcome of CE is the creation of a DWD system, where
the giant has become a HeWD, but in cases where the giant
core was non-degenerate a naked helium star (HeMS) is the
resulting companion.
\label{t:table3}
}
\tabletypesize\footnotesize
\tablehead{
$T_{\rm RLOF}$ & \multicolumn{2}{c}{Types} & $M_1$ & $M_2$ & $q$ &
${M'}_2$ & \multicolumn{1}{c}{Period} & Legend
}
\startdata
 222 & GB & CO  & 3.48 & 1.09 & 3.19 & 4.02 & $5.1941 \times 10^{1}$ & PRIM-CE(HeMS) \\
1630 & GB & CO  & 1.64 & 0.62 & 2.65 & 1.94 & $2.1042 \times 10^{1}$ & PRIM-PB-EXCH-CE \\
1639 & GB & CO  & 1.64 & 0.62 & 2.65 & 1.89 & $4.7514 \times 10^{1}$ & EXCH-CE \\
1893 & GB & CO  & 1.56 & 0.68 & 2.29 & 1.89 & $3.6108 \times 10^{1}$ & EXCH-CE \\
2969 & GB & CO  & 1.34 & 0.66 & 2.03 & 1.77 & $4.5509 \times 10^{0}$ & EXCH-CE \\
2598 & GB & CO  & 1.41 & 0.60 & 2.35 & 1.71 & $1.8143 \times 10^{1}$ & EXCH-CE \\
2450 & GB & CO  & 1.44 & 0.69 & 2.09 & 1.70 & $1.6049 \times 10^{2}$ & PRIM-CE \\
2116 & GB & CO  & 1.50 & 0.55 & 2.73 & 1.68 & $4.6434 \times 10^{1}$ & EXCH-CE \\
3229 & GB & CO  & 1.31 & 0.68 & 1.93 & 1.68 & $2.0417 \times 10^{1}$ & EXCH-CE \\
3303 & GB & CO  & 1.30 & 0.54 & 2.41 & 1.50 & $3.1223 \times 10^{1}$ & EXCH-CE \\
3860 & GB & CO  & 1.24 & 0.55 & 2.25 & 1.36 & $1.0945 \times 10^{2}$ & EXCH-CE \\
\tableline
 334 & HG & CO  & 4.07 & 0.57 & 7.14 & 3.99 & $2.3642 \times 10^{1}$ & PRIM-CE(HeMS) \\
1856 & GB & ONe & 1.57 & 1.01 & 1.55 & 2.33 & $6.2060 \times 10^{0}$ & PRIM-CE \\
1299 & GB & CO  & 1.77 & 0.88 & 2.01 & 2.28 & $5.3945 \times 10^{1}$ & PRIM-CE \\
1781 & HG & CO  & 1.57 & 0.56 & 2.80 & 1.92 & $1.9978 \times 10^{0}$ & EXCH-SMT \\
2153 & GB & CO  & 1.50 & 0.72 & 2.08 & 1.90 & $2.6303 \times 10^{1}$ & EXCH-CE \\
1707 & GB & CO  & 1.62 & 0.56 & 2.89 & 1.85 & $2.1828 \times 10^{1}$ & EXCH-CE \\
1858 & GB & CO  & 1.57 & 0.58 & 2.71 & 1.80 & $3.8619 \times 10^{1}$ & EXCH-CE \\
2301 & GB & CO  & 1.46 & 0.57 & 2.56 & 1.72 & $2.0297 \times 10^{1}$ & EXCH-CE \\
2672 & GB & He  & 1.40 & 0.47 & 2.98 & 1.53 & $3.1203 \times 10^{1}$ & EXCH-CE \\
2858 & GB & He  & 1.35 & 0.32 & 4.22 & 1.47 & $2.2487 \times 10^{0}$ & PRIM-CE \\
\tableline
 672 & GB & CO  & 2.43 & 0.77 & 3.16 & 2.88 & $6.1260 \times 10^{0}$ & EXCH-CE(HeMS) \\
4070 & GB & CO  & 1.36 & 0.71 & 1.92 & 1.74 & $4.6309 \times 10^{1}$ & PRIM-CE \\
2875 & GB & He  & 1.51 & 0.41 & 3.66 & 1.64 & $1.7278 \times 10^{1}$ & EXCH-CE \\
\tableline 
1226 & GB & CO  & 1.98 & 0.77 & 2.57 & 2.43 & $3.1687 \times 10^{1}$ & PRIM-CE \\
1597 & GB & CO  & 1.80 & 0.67 & 2.69 & 2.27 & $2.3700 \times 10^{0}$ & EXCH-CE \\ 
1486 & GB & CO  & 1.86 & 0.64 & 2.91 & 2.23 & $9.4690 \times 10^{0}$ & EXCH-CE \\
2192 & GB & CO  & 1.64 & 0.55 & 2.98 & 1.91 & $1.6321 \times 10^{1}$ & EXCH-CE \\
2564 & GB & CO  & 1.55 & 0.55 & 2.82 & 1.90 & $2.8284 \times 10^{0}$ & EXCH-CE \\
1783 & GB & He  & 1.74 & 0.35 & 4.97 & 1.87 & $3.5781 \times 10^{0}$ & EXCH-CE \\
2898 & GB & He  & 1.50 & 0.36 & 4.17 & 1.63 & $6.1525 \times 10^{0}$ & EXCH-CE \\
\enddata
\end{deluxetable}

\clearpage

\clearpage

\clearpage

\clearpage

\clearpage

\clearpage

\clearpage

\begin{thebibliography}{}
\bibitem[Aarseth, H\'{e}non \& Wielen (1974)]{aar74} Aarseth, S., 
   H\'{e}non, M., \& Wielen, R. 1974, \aap, 37, 183 
\bibitem[Aarseth (1999)]{aar99} Aarseth, S. J. 1999, \pasp, 111, 1333
\bibitem[Cappellaro et al. (1997)]{cap97} Cappellaro, E., Turatto, M., 
   Tsvetkov, D.Yu., Bartunov, O.S., Pollas, C., Evans, R., \& Hamuy, M. 
   1997, A\&A, 322, 431
\bibitem[Chen \& Leonard (1993)]{che93} Chen, K, \& Leonard, P.J.T. 1993, 
   \apj, 411, L75 
\bibitem[Chernoff \& Weinberg (1990)]{che90} Chernoff, D.F., \& Weinberg, M.D. 
   1990, \apj, 351, 121 
\bibitem[Duquennoy \& Mayor (1991)]{duq91} Duquennoy, A., \& Mayor, M. 
   1991, A\&A, 248, 485
\bibitem[Eggleton (2002)]{egg02} Eggleton, P.P. 2002, Evolutionary Processes in
   Binary and Multiple Stars, (Cambridge: Cambridge University Press),
   in preparation 
\bibitem[Eggleton, Fitchett \& Tout (1989)]{egg89} Eggleton, P.P., 
   Fitchett, M., \& Tout C.A. 1989, \apj, 347, 998 
\bibitem[Fan et al. (1996)]{fan96} Fan, X., et al. 1996, \aj, 112, 628
\bibitem[Giersz \& Heggie (1997)]{gie97} Giersz, M., \& Heggie, D.C. 1997,
   \mnras, 286, 709 
\bibitem[Guhathakurta et al. (1998)]{guh98} Guhathakurta, P., Webster, Z.T., 
   Yanny, B., Schneider, D.P., \& Bahcall, J.N. 1998, \aj, 116, 1757 
\bibitem[Heggie (1975)]{heg75} Heggie, D.C. 1975, \mnras, 173, 729
\bibitem[Heggie, Hut \& McMillan (1996)]{heg96} Heggie, D.C., Hut, P., \& 
   McMillan, S.L.W. 1996, \apj, 467, 359 
\bibitem[Heggie \& Rasio (1996)]{hnr96} Heggie, D.C., \& Rasio, F.A. 1996,
   \mnras, 282, 1064 
\bibitem[Hurley, Pols \& Tout (2000)]{hur00} Hurley, J.R., Pols, O.R., 
   \& Tout, C.A. 2000, \mnras, 315, 543
\bibitem[Hurley et al. (2001)]{hur01} Hurley, J. R., Tout, C. A.,
   Aarseth, S. J., \& Pols, O.R. 2001, \mnras, 323, 630
\bibitem[Hurley, Tout \& Pols (2002)]{hur02} Hurley, J.R., Tout, C.A., 
   \& Pols, O.R. 2002, \mnras, 329, 897 
\bibitem[Hurley \& Shara (2002)]{hsh02} Hurley, J.R., \& Shara, M.M. 2002, 
   \apj, in press
\bibitem[Iben \& Livio (1993)]{ibe93} Iben, I.-Jr., \& Livio, M. 1993, 
   \pasp, 105, 1373
\bibitem[Kraft (1983)]{kra83} Kraft, R. P. 1983, Highlights in Astronomy, 
   6, 129 
\bibitem[Kroupa, Tout \& Gilmore (1991)]{kro91} Kroupa, P., Tout, C. A., 
   \& Gilmore, G. 1991, \mnras, 251, 293 
\bibitem[Kroupa, Tout \& Gilmore (1993)]{kro93} Kroupa, P., Tout, C. A., 
   \& Gilmore, G. 1993, \mnras, 262, 545
\bibitem[Kurucz (1992)]{kur92} Kurucz R.L. 1992, in IAU Symposium 149, 
   The Stellar Populations of Galaxies, ed. B. Barbuy, \& A. Renzini  
   (Dordrecht: Kluwer), 225  
\bibitem[Lada, Strom \& Myers (1993)]{lad93} Lada, E. A., Strom, K. M., 
   \& Myers, P. C. 1993, in Protostars and Planets III,
   ed. E. Levy, \& J. Lunine (University of Arizona), 245
\bibitem[Leibundgut (2001)]{lei01} Leibundgut, B. 2001, \araa, 39, 67 
\bibitem[Makarov \& Fabricius (2001)]{mfa01} Makarov, V.V., 
   \& Fabricius, C. 2001, \aap, 368, 866 
\bibitem[Makino (2001)]{mak01} Makino, J. 2001, in ASP Conf. Ser. XX, 
   Stellar Collisions, Mergers and their Consequences, ed. M. M. Shara 
   (San Francisco: ASP), in press
\bibitem[Mardling \& Aarseth (2001)]{mar01} Mardling, R.A.,
   \& Aarseth, S.J. 2001, \mnras, 321, 398
\bibitem[Perlmutter et al. (1999)]{per99} Perlmutter, S., et al. 1999, 
   \apj, 517, 565
\bibitem[Pfahl, Rappaport \& Podsiadlowski (2002)]{pfa02} Pfahl, E., 
   Rappaport, S., \& Podsiadlowski, Ph. 2002, \apj, submitted 
\bibitem[Phillips (1993)]{phi93} Phillips, M.M. 1993, \apj, 413L, 105
\bibitem[Portegies Zwart \& Verbunt (1996)]{por96} Portegies Zwart, S.F.,
   \& Verbunt F. 1996, \aap, 309, 179 
\bibitem[Riess et al. (1998)]{rie98} Riess, A.G. et al. 1998, \aj, 116, 1009
\bibitem[Riess (2000)]{rie00} Riess, A.G. 2000, \pasp, 112, 1284
\bibitem[Saffer, Livio \& Yungelson (1998)]{saf98} Saffer, R.A., Livio, M., 
   \& Yungelson, L.R. 1998, \apj, 502, 394
\bibitem[Saio \& Nomoto (1998)]{sai98} Saio, H., \& Nomoto, K. 1998, 
   \apj, 500, 388 
\bibitem[Tout et al. (1997)]{tou97} Tout, C.A., Aarseth, S.J., Pols, O.R.,
   \& Eggleton P.P. 1997, \mnras, 291, 732 
\bibitem[Tout et al. (2001)]{tou01} Tout, C.A., Reg\"{o}s, E., 
   Wickramasinghe, D., Hurley, J.R., \& Pols, O.R. 2001, in 
   ASP Conf. Ser. 229, Evolution of Binary and Multiple Star Systems: 
   A Meeting in Celebration of Peter Eggleton's 60th Birthday, 
   ed. Ph. Podsiadlowski, S. Rappaport, A. R. King, F. D'Antona, \& L. Burder, 
   (San Francisco: ASP), 275 
\bibitem[Tutukov \& Yungelson (1994)]{tut94} Tutukov, A., \& Yungelson, L. 
   1994, \mnras, 268, 871 
\bibitem[Tutukov \& Yungelson (1996)]{tut96} Tutukov, A., \& Yungelson, L.
   1994, \mnras, 280, 1035 
\bibitem[Webbink (1988)]{web88} Webbink R.F. 1998, in IAU Colloquim 103, 
   The Symbiotic Phenomenon, ed. J. Mikolajewska, M. Friedjung, S.J. Kenyon, 
   \& R. Viotto (Dordrecht: Kluwer), 311 
\bibitem[Yungelson \& Livio (1998)]{yun98} Yungelson, L., \& Livio, M. 1998, 
   \apj, 497, 168 
\bibitem[Yungelson \& Livio (2000)]{yun00} Yungelson, L., \& Livio, M. 2000, 
   \apj, 528, 108
\bibitem[Yungelson et al. (1996)]{yun96} Yungelson, L., Livio, M., 
   Truran, J.W., Tutukov, A., \& Federova, A.V. 1996, \apj, 466, 890 
\end{thebibliography}
\end{document}